\documentclass[aps,pre,preprint,groupedaddress]{revtex4-1}%
\usepackage{graphicx}
\usepackage{tabularx}
\usepackage{dcolumn}
\usepackage{longtable}
\usepackage{tensor}
\usepackage{color}
\usepackage{placeins}
\usepackage{bm}
\usepackage{amsmath}
\usepackage{amsfonts}
\usepackage{amssymb}
\usepackage{float}%
\providecommand{\U}[1]{\protect\rule{.1in}{.1in}}
\begin{document}
\title{On-the-fly \emph{ab initio} three thawed Gaussians approximation: A
semiclassical approach to Herzberg-Teller spectra}
\author{Tomislav Begu\v{s}i\'{c}}
\author{Aur\'{e}lien Patoz}
\author{Miroslav \v{S}ulc}
\author{Ji\v{r}\'i Van\'i\v{c}ek}
\email{jiri.vanicek@epfl.ch.}
\affiliation{Laboratory of Theoretical Physical Chemistry, Institut des Sciences et
Ing\'enierie Chimiques, Ecole Polytechnique F\'ed\'erale de Lausanne (EPFL),
CH-1015, Lausanne, Switzerland}
\date{\today}

\begin{abstract}
Evaluation of symmetry-forbidden or weakly-allowed vibronic spectra requires
treating the transition dipole moment beyond the Condon approximation. To
include vibronic spectral features not captured in the global harmonic models,
we have recently implemented an on-the-fly \textit{ab initio} extended thawed
Gaussian approximation, where the propagated wavepacket is a Gaussian
multiplied by a linear polynomial. To include more anharmonic effects, here we
represent the initial wavepacket by a superposition of three independent
Gaussian wavepackets---one for the Condon term and two displaced Gaussians for
the Herzberg--Teller part. Application of this \textit{ab initio}
\textquotedblleft three thawed Gaussians approximation\textquotedblright\ to
vibrationally resolved electronic spectra of the phenyl radical and benzene
shows a clear improvement over the global harmonic and Condon approximations.
The orientational averaging of spectra, the relation between the gradient of
the transition dipole moment and nonadiabatic coupling vectors, and the
details of the extended and three thawed Gaussians approximation are discussed.

\end{abstract}
\maketitle

%\author[LCPT]{Tomislav Begu\v{s}i\'{c}}
%\author[LCPT]{Aur\'{e}lien Patoz}
%\author[LCPT]{Miroslav \v{S}ulc}
%\author[LCPT]{Ji\v{r}\'i Van\'i\v{c}ek\corref{cor1}}
%\ead{jiri.vanicek@epfl.ch}

%\address[LCPT]{Laboratory of Theoretical Physical Chemistry, Institut des Sciences et
%Ing\'enierie Chimiques, Ecole Polytechnique F\'ed\'erale de Lausanne (EPFL),
%CH-1015, Lausanne, Switzerland}

\graphicspath{{"d:/Group Vanicek/Desktop/Herzberg-Teller_papers/Herzberg-Teller_3Gauss/figures/"}{./figures/}{C:/Users/Jiri/Dropbox/Papers/Chemistry_papers/2018/Herzberg-Teller_3Gauss/revision_v4/revision_v4/}}

\section{\label{sec:intro}Introduction}

Electronic spectroscopy lies at the core of modern physical chemistry; not
only has it been the driving force in developing first insights into the
atomic and molecular structure \cite{Quack_Merkt:2011,Herzberg:1966}, but it
is also the method of choice for unraveling essential chemical and physical
processes. The time-dependent approach to electronic spectroscopy obtains the
spectrum as a Fourier transform of an appropriate time correlation function
\cite{Heller:1981a,Mukamel:1995}, which requires performing molecular quantum
dynamics, but provides more dynamical information about the interaction of the
molecule with light. Even though only relatively short (femtosecond to
picosecond) time scales are typically involved in electronic spectroscopy,
exact methods for solving the time-dependent Schr\"{o}dinger equation become
quickly impractical for larger molecules due to the exponential scaling with
the number of degrees of freedom \cite{Meyer_Worth:2009,Gatti:2014}. In
contrast, approximate semiclassical methods reduce the exponential quantum
problem to the propagation of classical trajectories, while often maintaining
sufficient accuracy, at least at short times. With the advent of
computationally feasible, though not necessarily cheap, \textit{ab initio}
calculations, semiclassical dynamics benefits from yet another advantage over
quantum dynamical methods: the formulation in terms of classical trajectories
allows for an efficient on-the-fly implementation, overcoming the need for
exploring the full potential energy surface.

Among many semiclassical approximations
\cite{Miller:1970,Miller:2001,Herman_Kluk:1984,Kay:2005,Zhang_Pollak:2004,Grossmann:2006,Mollica_Vanicek:2011,Ceotto_Conte:2017,Buchholz_Ceotto:2018}%
, one of the simplest, yet very intuitive and often surprisingly accurate, is
the thawed Gaussian approximation (TGA) of Heller \cite{Heller:1975}. Although
his work on Gaussian wavepackets \cite{Heller:1981b} inspired a number of
\textquotedblleft direct\textquotedblright, i.e., on-the-fly \emph{ab initio}
quantum
\cite{Ben-Nun_Martinez:2000,Saita_Shalashilin:2012,Makhov_Shalashilin:2017,Richings_Lasorne:2015,Curchod_Martinez:2018}
and semiclassical
\cite{Tatchen_Pollak:2009,Ceotto_Atahan:2009a,Ceotto_Atahan:2009b,Wong_Roy:2011,Ianconescu_Pollak:2013,Gabas_Ceotto:2017}
methods, the original thawed Gaussian approximation was only revisited
recently, when it was implemented within the on-the-fly \textit{ab initio}
framework for evaluating molecular absorption, emission, and photoelectron
spectra \cite{Wehrle_Vanicek:2014,Wehrle_Vanicek:2015}. Due to its
computational efficiency in comparison with exact quantum dynamics methods,
the TGA can account for \cite{Wehrle_Vanicek:2014,Wehrle_Vanicek:2015} the
full dimensionality of the system and mode-mixing (Duschinsky effect), both of
which can significantly influence the wavepacket dynamics
\cite{Mahapatra_Domcke:2005}. Recently, we have implemented
\cite{Patoz_Vanicek:2018} an extension to the on-the-fly \emph{ab initio}
thawed Gaussian approximation, aimed at describing the Herzberg--Teller
contribution to the vibrational structure of electronic absorption spectra.
The \emph{extended thawed Gaussian approximation} (ETGA)
\cite{Lee_Heller:1982,Patoz_Vanicek:2018}, propagates a Gaussian wavepacket
multiplied by a general polynomial in nuclear coordinates using a \emph{local
harmonic approximation}, i.e., the second-order expansion of the potential
about the center of the wavepacket. Such a wavepacket, however, is wider than
a simple Gaussian wavepacket, which raises the question whether more than a
single guiding classical trajectory should be used to correctly account for
the anharmonicity of the potential energy surface.

Here, we present the on-the-fly \textit{ab initio} implementation of an
alternative method for treating Herzberg--Teller spectra beyond the Condon
approximation. This semiclassical method is also based on the thawed Gaussian
approximation, yet, unlike the extended TGA, resolves the initial
Herzberg--Teller wavepacket into three well-defined Gaussians and propagates
them independently. This \textquotedblleft\emph{three thawed Gaussians
approximation}\textquotedblright\ (3TGA) is applied to evaluate the absorption
spectra of the phenyl radical and benzene. The 3TGA is compared with the
Condon approximation in order to analyze the importance of the
Herzberg--Teller contribution, with the global harmonic approaches to assess
the extent of anharmonicity, and with the extended TGA to analyze the effects
of wavepacket splitting.

\emph{Notation}: As the analysis of electronic spectra involves three
different vector spaces, let us summarize the notation for reference here,
although most should be clear from the context. If $D$ denotes the number of
nuclear degrees of freedom and $S$ the number of electronic states considered,
the three vector spaces are the nuclear $D$-dimensional real coordinate space
$%
%TCIMACRO{\U{211d} }%
%BeginExpansion
\mathbb{R}
%EndExpansion
^{D}$, electronic $S$-dimensional complex Hilbert space $%
%TCIMACRO{\U{2102} }%
%BeginExpansion
\mathbb{C}
%EndExpansion
^{S}$, and the ambient $3$-dimensional space $%
%TCIMACRO{\U{211d} }%
%BeginExpansion
\mathbb{R}
%EndExpansion
^{3}$. To distinguish these spaces, $3$-dimensional vectors will be denoted
with an arrow (e.g., $\vec{\epsilon}$), whereas $D$-dimensional nuclear
vectors or $D\times D$ matrices will use no special notation (e.g.,
generalized nuclear coordinates $q$). Scalar and matrix products in both the
$3$-dimensional and $D$-dimensional nuclear spaces will be denoted with a dot
(as in $\vec{\mu}_{21}\cdot\vec{\epsilon}$ or $p^{T}\cdot m^{-1}\cdot p$). We
shall use the \textbf{bold} font for $S\times S$ matrices (such as $\bm{\mu}$)
representing electronic operators expressed in the $S$-state basis of the
electronic Hilbert space. The matrix product will use no special notation; it
will be expressed by a juxtaposition of the matrices (as in $\mathbf{AB}$).

\section{\label{sec:theory}Theory}

\subsection{\label{subsec:absspec}Time-dependent approach to spectroscopy}

Within the electric-dipole approximation and first-order time-dependent
perturbation theory, the absorption cross section for a linearly polarized
light of frequency $\omega$ can be expressed as the Fourier transform
\begin{equation}
\sigma(\vec{\epsilon},\omega)=\frac{2\pi\omega}{\hbar c}\int_{-\infty}%
^{\infty}C_{\mu\mu}(\vec{\epsilon},t)e^{i\omega t}dt \label{eq:sigma}%
\end{equation}
of the dipole time autocorrelation function $C_{\mu\mu}(\vec{\epsilon},t)$.
Assuming the zero temperature approximation, the initial state is
$|1,g\rangle$, i.e., the ground vibrational state $g$ of the ground electronic
state $1$. Assuming, furthermore, that the incident radiation is in resonance
only with a single pair of electronic states $1$ and $2$ that are not
vibronically coupled, the dipole time autocorrelation function reduces to
\cite{Heller:1981a}
\begin{equation}
C_{\mu\mu}(\vec{\epsilon},t)=\langle1,g|e^{i\hat{H}_{1}t/\hbar}\hat{\mu}%
_{12}e^{-i\hat{H}_{2}t/\hbar}\hat{\mu}_{21}|1,g\rangle\,, \label{eq:C_mm}%
\end{equation}
where $\hat{H}_{1}$ and $\hat{H}_{2}$ are nuclear Hamiltonian operators in the
ground and excited electronic states, and
\begin{equation}
\hat{\mu}_{21}:=\hat{\vec{\mu}}_{21}\cdot\vec{\epsilon} \label{eq:mu_21}%
\end{equation}
is the projection of the matrix element $\hat{\vec{\mu}}_{21}$ of the
molecular transition dipole moment matrix $\hat{\vec{\bm{\mu}}}$ on a
three-dimensional polarization unit vector $\vec{\epsilon}$ of the electric field.

\subsection{\label{subsec:orient_averaging}Orientational averaging of the
spectrum}

In gas phase or another isotropic medium, one has to average the vibronic
spectrum (\ref{eq:sigma}) over all orientations of the molecule with respect
to the polarization $\vec{\epsilon}$ of the electric field. Within the Condon
approximation, the transition dipole moment is independent of coordinates, and
this averaging is trivial, but for a general dipole moment, one has to be more
careful. It is surprising that in theoretical papers on vibronic spectroscopy,
the averaging is often ignored despite an extensive work on orientational
averaging of both linear and nonlinear spectra
\cite{Andrews_Thirunamachandran:1977,Gelin_Domcke:2017}.

It is useful to define the spectrum \emph{tensor} $\overleftrightarrow{\sigma
}(\omega)$, from which the spectrum (\ref{eq:sigma}) for a specific
polarization $\vec{\epsilon}$, is obtained by \textquotedblleft
evaluation\textquotedblright:%
\begin{equation}
\sigma(\vec{\epsilon},\omega)=\vec{\epsilon}^{\,T}\cdot
\overleftrightarrow{\sigma}(\omega)\cdot\vec{\epsilon}.
\label{eq:sigma_epsilon}%
\end{equation}
The \emph{orientational averaging} of the spectrum corresponds to the
averaging of $\sigma(\vec{\epsilon},\omega)$ over all unit vectors
$\vec{\epsilon}$. Remarkably, due to the isotropy of the 3-dimensional
Euclidean space, the average over all orientations does not have to be
performed numerically but can be reduced to a simple arithmetic average over
only three arbitrary orthogonal orientations of the molecule with respect to
the field:%
\begin{align}
\overline{\sigma(\vec{\epsilon},\omega)}  &  =\frac{1}{3}\operatorname*{Tr}%
\overleftrightarrow{\sigma}(\omega)\nonumber\\
&  =\frac{1}{3}\left[  \sigma_{xx}(\omega)+\sigma_{yy}(\omega)+\sigma
_{zz}(\omega)\right] \label{eq:sigma_average}\\
&  =\frac{1}{3}\left[  \sigma(\vec{e}_{x},\omega)+\sigma(\vec{e}_{y}%
,\omega)+\sigma(\vec{e}_{z},\omega)\right]  .\nonumber
\end{align}
Appendix~\ref{sec_app:orient_av} contains an explicit proof of these
equalities, of which the first is coordinate-independent and the other two are
explicit in Cartesian coordinates ($\vec{e}_{x}$ denotes the unit vector along
the $x$-axis.). This final result for the orientational average is exact for
any coordinate dependence of the transition dipole operator and also for any
quantum or semiclassical dynamical method for evaluating the autocorrelation
function. Moreover, it is valid even for arbitrary pure or mixed initial
molecular states and arbitrary nonadiabatic or electric-dipole couplings among
the $S$ electronic states. As the Fourier transform is a linear operation, one
may choose to apply the averaging already to the dipole autocorrelation
function instead of the spectrum:%
\begin{equation}
\overline{C_{\mu\mu}(\vec{\epsilon},t)}=\frac{1}{3}\operatorname{Tr}%
[\overleftrightarrow{C}_{\mu\mu}(t)]. \label{eq:C_average}%
\end{equation}

Now let us go back to our two-state system, where only one element of the
electric transition dipole moment, $\vec{\mu}_{21}(q)$ plays a role and
$C_{\mu\mu}(\vec{\epsilon},t)$ is given by Eq.~(\ref{eq:C_mm}). Then it is
convenient to suppress the subscripts $21$ and denote the $21$ element simply
as $\vec{\mu}(q)$, which we shall do for the remainder of this subsection.
Among a continuum of other possibilities, expression (\ref{eq:C_average})
provides two simple yet exact recipes for the orientational average: one can
take the arithmetic average of $C_{\mu\mu}(\vec{\epsilon},t)$ either for three
orthogonal polarizations $\vec{\epsilon}$ and a fixed molecular orientation
\cite{Hein_Rodriguez:2012},
\begin{equation}
\overline{C_{\mu\mu}(\vec{\epsilon},t)}=\frac{1}{3}\left[  C_{\mu_{x}\mu_{x}%
}(t)+C_{\mu_{y}\mu_{y}}(t)+C_{\mu_{z}\mu_{z}}(t)\right]  ,
\label{eq:C_average_xyz}%
\end{equation}
or for three orthogonal orientations of the molecule and a fixed polarization
$\vec{\epsilon}$. Within the Condon approximation, in which $\vec{\mu}$ is
coordinate-independent, the orientational average (\ref{eq:C_average_xyz})
simplifies further into the standard textbook recipe%
\begin{equation}
\overline{C_{\mu\mu}(\vec{\epsilon},t)}=\frac{1}{3}C_{|\vec{\mu}||\vec{\mu}%
|}(t), \label{eq:C_average_CA}%
\end{equation}
where $|\vec{\mu}|$ is the magnitude of the transition dipole moment, so only
a single calculation is required---for the transition dipole moment aligned
with the field.

\subsection{\label{subsec:CA_HT}Condon and Herzberg--Teller approximations for
the transition dipole moment}

The electric transition dipole moment is, in general, a function of nuclear
coordinates, yet, within the \emph{Condon approximation} \cite{Condon:1926},
this moment is assumed to be independent of the molecular geometry and is
approximated to the zeroth order around the initial geometry:
\begin{equation}
\vec{\bm{\mu}}(q)\approx\vec{\bm{\mu}}(q_{\text{eq}})\,.\label{eq:CA}%
\end{equation}
The widespread use of this approximation is justified by its validity in many
systems, as it can describe most of the strongly symmetry-allowed transitions
both qualitatively and quantitatively. However, a number of molecules exhibit
symmetry-forbidden (also called \textquotedblleft electronically
forbidden\textquotedblright) transitions, i.e., transitions $\alpha
\leftarrow\beta$ with $\vec{\mu}_{\alpha\beta}(q_{\text{eq}})=0$, which cannot
be described within the Condon approximation. Such systems, as well as systems
in which the Condon term is small but not exactly zero, can be treated with
the \emph{Herzberg--Teller approximation} \cite{Herzberg_Teller:1933} that
takes into account also the gradient of the transition dipole moment with
respect to nuclear degrees of freedom:
\begin{equation}
\vec{\bm{\mu}}(q)\approx\vec{\bm{\mu}}(q_{\text{eq}})+\partial_{q}%
\vec{\bm{\mu}}|_{q_{\text{eq}}}^{T}\cdot(q-q_{\text{eq}})\,,\label{eq:HTA}%
\end{equation}
where $\partial_{q}$ is an abbreviation for $\partial/\partial q$.

\subsection{\label{subsec:grad_mu_NAC}Relationship between the gradient of the
transition dipole moment and nonadiabatic vector couplings}

Although the Herzberg--Teller approximation is often discussed in terms of
vibronic couplings between different electronic states
\cite{Seidner_Domcke:1992}, this relation is not obvious from
Eq.~(\ref{eq:HTA}). Here we demonstrate this relationship explicitly. In
particular, we shall prove a remarkable equality%
\begin{equation}
\partial_{j}\vec{\bm{\mu}}(q)=\left[  \vec{\bm{\mu}}(q),\mathbf{F}%
_{j}(q)\right]  +\partial_{j}\vec{\mu}_{\text{nu}}(q)\mathbf{1}
\label{eq:grad_mu}%
\end{equation}
satisfied by the gradient of the matrix representation $\vec{\bm{\mu}}(q)$ of
the molecular dipole operator $\hat{\vec{\mu}}_{\text{mol}}$ at the nuclear
configuration $q$. In this equation, matrix elements of $\vec{\bm{\mu}}(q)$
are defined as partial, electronic expectation values%
\begin{equation}
\vec{\mu}_{\alpha\beta}(q):=\langle\alpha(q)|\hat{\vec{\mu}}_{\text{mol}%
}|\beta(q)\rangle, \label{eq:mu_ab}%
\end{equation}
elements of the matrix $\mathbf{F}_{j}(q)$ of nonadiabatic vector couplings
are obtained as%
\begin{equation}
F_{\alpha\beta,j}:=\langle\alpha(q)|\partial_{j}\beta(q)\rangle,
\label{eq:F_ab_j}%
\end{equation}
and $\vec{\mu}_{\text{nu}}$ is the nuclear component of $\hat{\vec{\mu}%
}_{\text{mol}}$:%
\begin{equation}
\vec{\mu}_{\text{nu}}(q)=e\sum_{N=1}^{N_{\text{nu}}}Z_{N}\vec{R}_{N}(q).
\label{eq:mu_nu}%
\end{equation}
Here $Z_{N}$ is the atomic number and $\vec{R}_{N}(q)$ the Cartesian
coordinates of the $N$th nucleus expressed as a function of the generalized
nuclear coordinates $q$, which can be Cartesian, normal-mode, Jacobi, or any
other coordinates.

Direct interpretation of relation (\ref{eq:grad_mu}), which is proven in
Appendix~\ref{sec_app:grad_mu_NAC}, explains the concepts of \emph{vibronic
transitions} and \emph{intensity borrowing}. Namely, the gradient of the
transition dipole moment between states $\alpha$ and $\beta$ can be nonzero
only if there exists an intermediate state $\gamma$ that is nonadiabatically
(i.e., \emph{vibronically}) coupled to one of the states and electric-dipole
coupled to the other state. This is even more explicit in the component
expression (\ref{eq:grad_mu_components}) in the penultimate line of the proof
in the Appendix. Typically, the nonadiabatic couplings with the ground state
($\beta=1$) can be neglected at the ground-state optimized geometry, around
which the transition dipole moment is expanded. This leads to an expression
\cite{Li_Lin:2010}
\begin{equation}
\partial_{j}\vec{\mu}_{\alpha1}\approx-\sum_{\mathbf{\gamma}}F_{\alpha
\gamma,j}\vec{\mu}_{\gamma1} \label{eq:vibronic_coupling_final}%
\end{equation}
that explains the meaning of \emph{intensity borrowing}, in which the symmetry
forbidden transition to the formally dark state $\alpha$ occurs due to a
borrowing of the transition intensity from the neighboring bright electronic
states $\gamma$ that are vibronically coupled to the state $\alpha$. Note
that, despite introducing nonadiabatic couplings between excited electronic
states, the original Born-Oppenheimer picture adopted in
Subsection~\ref{subsec:absspec} remains valid; the vibronic couplings that
induce the transition do not necessarily influence the nuclear wavepacket
dynamics. Although these nonadiabatic couplings are essential for describing
the existence of symmetry-forbidden spectra, they are still rather small and
their contribution to the field-free Hamiltonian of the system is negligible.
The rather high resolution of the absorption spectrum of benzene supports
these considerations---otherwise, significant population transfer would lead
to the shortening of the excited-state lifetime and, consequently, to
significant broadening of the spectral lines.

\subsection{\label{subsec:ETGA}Extended thawed Gaussian approximation}

To evaluate the dipole time autocorrelation function (\ref{eq:C_mm}%
)\ corresponding to a specific polarization of the electric field, let us
rewrite it as
\begin{equation}
C_{\mu\mu}(\vec{\epsilon},t)=C(t)e^{iE_{1,g}t/\hbar} \label{eq:C_mm__to_C_t}%
\end{equation}
in terms of the vibrational zero-point energy $E_{1,g}$ of the ground
electronic state and the wavepacket autocorrelation function
\begin{equation}
C(t)=\langle\phi(0)|\phi(t)\rangle\, \label{eq:C_t}%
\end{equation}
for the initial wavepacket $|\phi(0)\rangle=\hat{\mu}_{21}|1,g\rangle$
propagated on the excited-state potential energy surface with the Hamiltonian
$\hat{H}_{2}$:
\begin{equation}
|\phi(t)\rangle=e^{-i\hat{H}_{2}t/\hbar}|\phi(0)\rangle. \label{eq:phi_t}%
\end{equation}

This propagation is the most demanding part of the spectra calculation and we
shall use an on-the-fly \textit{ab initio} semiclassical method for this
purpose. One of the simplest semiclassical methods, the TGA \cite{Heller:1975}%
, relies on the fact that a Gaussian wavepacket conserves its form when
propagated in a time-dependent harmonic potential; it extends the global
harmonic methods by approximating the potential energy surface locally to the
second order, in what is known as the \emph{local harmonic approximation}.
Although approximating the initial state in electronic spectroscopy with a
Gaussian is only reasonable within the Condon approximation, let us first
discuss this simplest case because it will serve as a starting point for
extensions to more general forms of the initial wavepacket needed to describe
Herzberg--Teller spectra.

Gaussian wavepacket considered in TGA is given in the position representation
as
\begin{equation}
\psi_{t}(q)=N_{0}\exp{\left\{  -(q-q_{t})^{T}\cdot A_{t}\cdot(q-q_{t}%
)+\frac{i}{\hbar}\left[  p_{t}^{T}\cdot(q-q_{t})+\gamma_{t}\right]  \right\}
} \label{eq:GWP}%
\end{equation}
where $N_{0}$ is a normalization constant, $(q_{t},p_{t})$ are the phase-space
coordinates of the center of the wavepacket, $A_{t}$ is a symmetric complex
width matrix, and $\gamma_{t}$ a complex number whose real part is a dynamical
phase and imaginary part guarantees the normalization for times $t>0$. The
wavepacket is propagated with a Hamiltonian
\begin{equation}
\hat{H}_{\text{eff}}(t)\equiv H_{\text{eff}}(\hat{q},\hat{p},t)=\frac{1}%
{2}\hat{p}^{T}\cdot m^{-1}\cdot\hat{p}+V_{\text{eff}}(\hat{q},t),
\label{eq:effHam}%
\end{equation}
where $m$ is the diagonal mass matrix and $V_{\text{eff}}$ an effective
time-dependent potential given by the local harmonic approximation of the true
potential $V$ at the center of the wavepacket:
\begin{equation}
V_{\text{eff}}(q,t)=V|_{q_{t}}+(\partial_{q}V|_{q_{t}})^{T}\cdot
(q-q_{t})+\frac{1}{2}(q-q_{t})^{T}\cdot\operatorname{Hess}_{q}V|_{q_{t}}%
\cdot(q-q_{t})\,. \label{eq:effPot}%
\end{equation}
To avoid the confusion between the Laplacian and Hessian operators, both of
which are often denoted by $\partial^{2}/\partial q^{2}$, we use the notation
$\operatorname{Hess}_{q}V$ for Hessian matrix with respect to nuclear
coordinates:%
\[
\left(  \operatorname{Hess}_{q}V\right)  _{ij}:=\partial_{i}\partial_{j}V.
\]
Insertion of the wavepacket ansatz (\ref{eq:GWP}) and the effective potential
(\ref{eq:effPot}) into the time-dependent Schr\"{o}dinger equation yields a
system of ordinary differential equations \cite{Heller:1975}
\begin{align}
\dot{q}_{t}  &  =m^{-1}\cdot p_{t}\,,\label{eq:q_dot}\\
\dot{p}_{t}  &  =-\partial_{q}\,V|_{q_{t}}\,,\label{eq:p_dot}\\
\dot{A}_{t}  &  =-2i\hbar A_{t}\cdot m^{-1}\cdot A_{t}+\frac{i}{2\hbar
}\text{$\operatorname{Hess}$}_{q}V|_{q_{t}}\,,\label{eq:A_dot}\\
\dot{\gamma}_{t}  &  =L_{t}-\hbar^{2}\operatorname{Tr}\left(  m^{-1}\cdot
A_{t}\right)  \, \label{eq:TGA_diff_expressions}%
\end{align}
for evolving parameters of the Gaussian; $L_{t}$ denotes the Lagrangian. This
system of differential equations, which is within the local harmonic
approximation equivalent to the solution of the Schr\"{o}dinger equation,
implies that phase-space coordinates $q_{t}$ and $p_{t}$ follow classical
equations of motion, while the propagation of the width $A_{t}$ and phase
$\gamma_{t}$ involves propagating the $2D\times2D$ stability matrix
\begin{equation}
M_{t}=%
\begin{pmatrix}
M_{t,qq} & M_{t,qp}\\
M_{t,pq} & M_{t,pp}%
\end{pmatrix}
:=%
\begin{pmatrix}
\frac{\partial q_{t}}{\partial q_{0}} & \frac{\partial q_{t}}{\partial p_{0}%
}\\
\frac{\partial p_{t}}{\partial q_{0}} & \frac{\partial p_{t}}{\partial p_{0}}%
\end{pmatrix}
, \label{eq:M_t}%
\end{equation}
which depends on the Hessians of the potential energy surface $V$. See
Appendix~\ref{sec_app:A_gamma_prop} for details.

To tackle a more general coordinate dependence of the transition dipole
moment, such as the Herzberg--Teller approximation (\ref{eq:HTA}), Lee and
Heller \cite{Lee_Heller:1982} proposed the \emph{extended TGA} (ETGA) that
considers a more general form of the initial wavepacket, namely a Gaussian
multiplied by a polynomial \cite{Lee_Heller:1982, Patoz_Vanicek:2018}:
\begin{equation}
\phi_{0}(q)=P\left(  q-q_{0}\right)  \psi_{0}(q).\label{eq:phi_t_q_polynomial}%
\end{equation}
Otherwise, the ETGA is the same as the original TGA; it uses the local
harmonic approximation (\ref{eq:effPot}) for the potential along the
trajectory $q_{t}$, but makes no other approximation.

Because the only dependence of $\psi_{0}(q)$ on $p_{0}$ comes from the
exponent $p_{0}^{T}\cdot\left(  q-q_{0}\right)  $ [see Eq.~(\ref{eq:GWP})],
the polynomial prefactor in Eq.~(\ref{eq:phi_t_q_polynomial}) can be replaced
by the \emph{same} polynomial in the derivatives with respect to $p_{0}$:
\begin{equation}
\phi_{0}(q)=P\left(  \frac{\hbar}{i}\frac{\partial}{\partial p_{0}}\right)
\psi_{0}(q). \label{eq:phi_0_p_polynomial}%
\end{equation}
In Appendix~\ref{sec_app:ETGA}, we prove that the local harmonic approximation
implies that the ETGA wavepacket at time $t$ has the simple form
\cite{Lee_Heller:1982}
\begin{equation}
\phi_{t}(q)=P\left(  \frac{\hbar}{i}\frac{\partial}{\partial p_{0}}\right)
\psi_{t}(q), \label{eq:phi_t_p_polynomial}%
\end{equation}
where the dependence of $\psi_{t}(q)$ on initial conditions $q_{0}$ and
$p_{0}$ is taken into account. In particular, equations of
motion~(\ref{eq:q_dot})-(\ref{eq:TGA_diff_expressions}) for $q_{t}$, $p_{t}$,
$A_{t}$, and $\gamma_{t}$ remain unchanged.

As for the parameters of the polynomial, here we consider only the constant
and linear terms required in the Herzberg--Teller approximation~(\ref{eq:HTA}%
). In Appendix~\ref{sec_app:ETGA}, we prove that \cite{Lee_Heller:1982}
\begin{equation}
\phi_{t}(q)=\left[  \mu_{21}(q_{0})+b_{t}^{T}\cdot(q-q_{t})\right]  \psi
_{t}(q), \label{eq:phi_t_HTA}%
\end{equation}
where the linear parameter of the polynomial at time $t$ is
\begin{equation}
b_{t}=\left(  -2i\hbar A_{t}\cdot M_{qp}+M_{pp}\right)  \cdot\partial_{q}%
\mu_{21}|_{q_{0}}\,. \label{eq:polynomial_eq_of_motion}%
\end{equation}
Because all ingredients needed in Eq.~(\ref{eq:polynomial_eq_of_motion}) are
already evaluated for the propagation of the parameters of the Gaussian, the
evaluation of $b_{t}$ comes at almost no additional cost.

\subsection{\label{subsec:3TGA}Three thawed Gaussians approximation}

\begin{figure}
\includegraphics[scale=1]{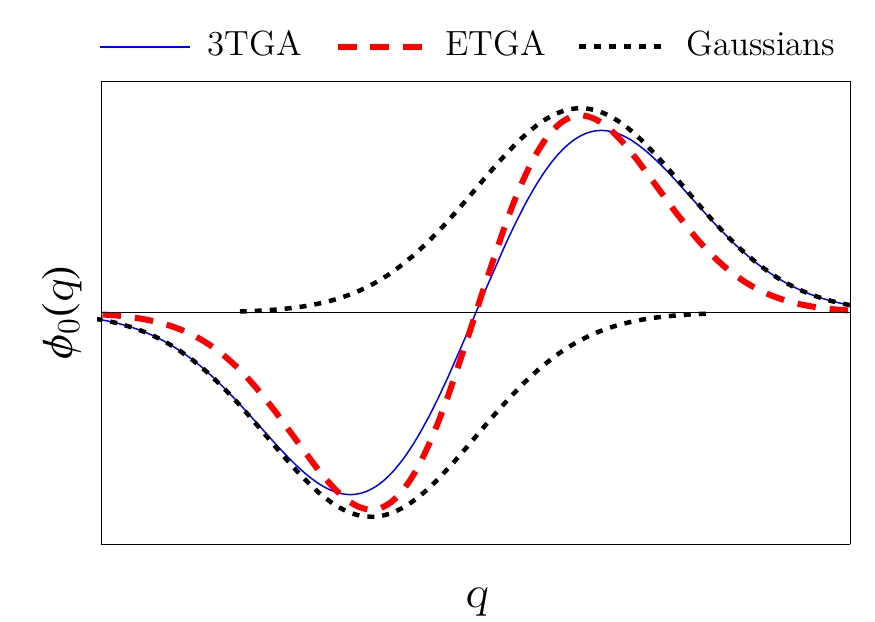}
\caption{\label{fig:3TGA_Scheme} The 3TGA wavepacket is obtained by replacing the extended TGA (ETGA)
wavepacket, i.e., a Gaussian multiplied by
a linear polynomial, by a sum of two displaced Gaussians (black dotted line).}
\end{figure}

Let us now describe another generalization of the TGA, which, as the ETGA, can
also evaluate the Herzberg--Teller spectra, but, in contrast to the ETGA, can
also account for wavepacket splitting. Recall that the ETGA wavepacket is
guided by a single trajectory propagated using the local harmonic
approximation of the potential at the center of the wavepacket. Whereas the
center of a Gaussian maximizes the probability density, this is false for
purely Herzberg--Teller wavepackets [Eq.~(\ref{eq:phi_t_HTA}) with $\mu
_{21}(q_{0})=0$], where the probability density at the center is zero.
Although the ETGA is still an exact solution of the time-dependent
Schr\"{o}dinger equation in a global harmonic potential, its performance in a
general potential can, therefore, be questioned \cite{Lee_Heller:1982}. To
further investigate the errors introduced by the local harmonic approximation,
we propose to approximate the Herzberg--Teller part
\begin{equation}
\phi_{0}^{\text{ETGA-HT}}(q)=\text{$\partial_{q}\mu_{21}$}^{T}\cdot
(q-q_{0})\,g_{q_{0}}(q)\, \label{eq:initial_HT}%
\end{equation}
of the initial state used in the ETGA\ with an antisymmetric linear
combination of two displaced Gaussians (see Fig.~\ref{fig:3TGA_Scheme}):
\begin{equation}
\phi_{0}^{\text{3TGA-HT}}(q)=f_{d}\,\left[  g_{q_{0}+\Delta_{d}}%
(q)-g_{q_{0}-\Delta_{d}}(q)\right]  \,, \label{eq:initial_3TGA-HT}%
\end{equation}
where $g_{q_{c}}(q)$ is a normalized Gaussian $\psi_{0}(q)$ centered at
$q_{c}$ instead of $q_{0}$, with zero initial momentum and phase
($p_{0}=\gamma_{0}=0$):
\begin{equation}
g_{q_{c}}(q):=N_{0}\,\,e^{-(q-q_{c})^{T}\cdot A_{0}\cdot(q-q_{c})}\,,
\label{eq:gaussian}%
\end{equation}
$\Delta_{d}$ is a displacement vector and $f_{d}$ a scaling factor. Note also
that the same width matrix $A_{0}$ is used for $g_{q_{c}}$ as for $\psi
_{0}(q)$.

We will now show that the approximative wavepacket $\phi_{0}^{\text{3TGA-HT}}$
converges to the exact Herzberg--Teller wavepacket $\phi_{0}^{\text{ETGA-HT}}%
$when the displacement $\Delta_{d}$ approaches zero along an appropriate
direction and a corresponding scaling factor $f_{d}$ is used. For brevity,
throughout this subsection we shall again write $\mu$ instead of $\vec{\mu
}_{21}(q_{0})\cdot\vec{\epsilon}$ and $\partial_{q}\mu$ instead of
$\partial_{q}(\vec{\mu}_{21}(q)\cdot\vec{\epsilon})|_{q_{0}}$.

First, let us discuss the shape of the exact wavepacket $\phi_{0}%
^{\text{ETGA-HT}}$. Obviously, in one nuclear dimension, $\phi_{0}%
^{\text{ETGA-HT}}$ has exactly two local extrema: a minimum and a maximum. In
Appendix \ref{sec_app:3TGA_extrema} we prove that this property holds for any
number $D$ of nuclear degrees of freedom:\ there are two and only two local
extrema, a maximum and minimum, located at%
\begin{equation}
q_{\text{max, min}}^{\text{ETGA-HT}}=q_{0}\pm\Delta q\,, \label{eq:max_HT}%
\end{equation}
with the displacement vector
\begin{equation}
\Delta q=\frac{A_{0}^{-1}\cdot\partial_{q}\mu}{\sqrt{2\partial_{q}\mu^{T}\cdot
A_{0}^{-1}\cdot\partial_{q}\mu}}. \label{eq:Delta_q}%
\end{equation}

It is rather obvious that in order for $\phi_{0}^{\text{3TGA-HT}}$ to be a
good fit of $\phi_{0}^{\text{ETGA-HT}}$, the two Gaussians in
Eq.~(\ref{eq:initial_3TGA-HT}) must be displaced from $q_{0}$ along the
direction of the extrema of $\phi_{0}^{\text{ETGA-HT}}$. We therefore choose
\begin{equation}
\Delta_{d}=d\,\Delta q, \label{eq:qd}%
\end{equation}
where the dimensionless parameter $d$ controls the magnitude of the
displacement. The scaling factor $f_{d}$ is obtained by equating the norms of
$\phi_{0}^{\text{3TGA-HT}}$ and $\phi_{0}^{\text{ETGA-HT}}$, which gives
\begin{equation}
f_{d}=\frac{\sqrt{2}}{4}\sqrt{\frac{\partial_{q}\mu^{T}\cdot A_{0}^{-1}%
\cdot\partial_{q}\mu}{1-e^{-d^{2}}}}. \label{eq:f}%
\end{equation}

A quick inspection reveals $\phi_{0}^{\text{ETGA-HT}}$ to be a directional
derivative of $g_{q_{0}}(q)$ and $\phi_{0}^{\text{3TGA-HT}}$ a
finite-difference approximation of this derivative, exact in the limit
$d\rightarrow0$. One can see this explicitly by noting that the normalized
initial overlap of the two wavepackets,
\begin{equation}
\frac{\langle\phi_{0}^{\text{ETGA-HT}}|\phi_{0}^{\text{3TGA-HT}}\rangle
}{\lVert\phi_{0}^{\text{ETGA-HT}}\rVert^{2}}=\frac{d\,e^{-d^{2}/4}}%
{\sqrt{1-e^{-d^{2}}}}, \label{eq:init_overlap}%
\end{equation}
is maximized and converges to unity as $d$ approaches zero. Either proof
implies that a better description of the initial wavepacket is obtained with
smaller values of $d$. On the other hand, anharmonicity effects are included
only when the two initial wavepackets are significantly displaced, implying
that a larger $d$ is needed. A natural displacement that is neither too small
nor too large corresponds to placing the two Gaussians at the extrema of the
wavepacket by setting $d=1$ and $\Delta_{d}=\Delta q$; in this case, the
initial overlap Eq.~(\ref{eq:init_overlap}) is $\approx98\%$. As this choice
of $d$ results in a large initial overlap and in trajectories that correspond
to the maxima of the probability density of the Herzberg--Teller component, we
recommend using $d=1$ in all applications. However, in the calculations
presented in Section~\ref{sec:resanddisc}, we will test not only $d=1$ but
also several smaller displacements, giving even larger initial overlaps.

Finally, by adding the Condon term, one obtains a superposition of three
Gaussian wavepackets that are propagated independently; the total initial
wavepacket in the \textquotedblleft three thawed Gaussians
approximation\textquotedblright\ (3TGA) is
\begin{equation}
\phi_{0}^{\text{3TGA}}(q)=\mu\,g_{q_{0}}(q)+f_{d}\,\left[  g_{q_{0}+\Delta
_{d}}(q)-g_{q_{0}-\Delta_{d}}(q)\right]  \,. \label{eq:3TGA}%
\end{equation}
In contrast to both the original and extended TGA, where the evolution given
by the time-dependent effective Hamiltonian is exactly unitary, the norm of
the 3TGA wavepacket is not conserved; since the three Gaussians are propagated
with three different effective Hamiltonians, their overlaps are time-dependent
and change the norm of the total 3TGA wavepacket. To see this, consider
several wavepackets $|\psi_{i}\rangle$, each propagated with its own evolution
operator $\hat{U}_{i}$. If $\hat{U}_{i}\neq\hat{U}_{j}$, then the overlap of
two such wavepackets is time-dependent:
\begin{equation}
\langle\psi_{i}(t)|\psi_{j}(t)\rangle=\langle\psi_{i}(0)|\hat{U}_{i}^{\dagger
}\hat{U}_{j}|\psi_{j}(0)\rangle\neq\langle\psi_{i}(0)|\psi_{j}(0)\rangle\,.
\label{eq:time_dep_overlap}%
\end{equation}
The time dependence of the norm of the 3TGA wavepacket arises from these
time-dependent overlaps since
\begin{align}
\left\Vert \phi^{\text{3TGA}}(t)\right\Vert ^{2}  &  =\langle\phi
^{\text{3TGA}}(t)|\phi^{\text{3TGA}}(t)\rangle\nonumber\\
&  =\mu^{2}+2f_{d}^{2}\left[  1-\text{Re}\left(  \langle g_{q_{0}+\Delta_{d}%
}^{t}|g_{q_{0}-\Delta_{d}}^{t}\rangle\right)  \right] \nonumber\\
&  +2\mu f_{d}\left[  \text{Re}\left(  \langle g_{q_{0}}^{t}|g_{q_{0}%
+\Delta_{d}}^{t}\rangle\right)  -\text{Re}\left(  \langle g_{q_{0}}%
^{t}|g_{q_{0}-\Delta_{d}}^{t}\rangle\right)  \right]  \,. \label{eq:3GNorm}%
\end{align}
Due to the time dependence of the norm, the calculated dipole time
autocorrelation function has to be renormalized at each time step.

\subsection{\label{subsec:OTF}\textit{Ab initio} implementation}

The \textit{ab initio} implementation of the thawed Gaussian approximation has
been discussed in Refs. \citenum{Wehrle_Vanicek:2014,Wehrle_Vanicek:2015};
only a brief overview is given here. The propagation of a thawed Gaussian
wavepacket requires a single classical trajectory on the excited-state
potential energy surface and the Hessians of the excited-state potential along
this trajectory. These data are evaluated in Cartesian coordinates and are
then transformed to mass-scaled normal mode coordinates, after removing the
translational degrees of freedom by translation to the center of mass frame
and rotational degrees of freedom by rotation to the Eckart frame. The choice
of the coordinates and correct coordinate transformation is the essence of the
on-the-fly \textit{ab initio} implementation; mass-scaled normal modes of the
ground electronic state are useful not only for a straightforward construction
of the initial wavepacket, but also for further interpretation of the results.

Within the 3TGA, each thawed Gaussian is propagated using the above mentioned
scheme. The initial positions are obtained in the following way: First, the
gradient of the transition dipole moment in Cartesian coordinates is projected
onto one of three orthogonal polarizations of the electric field and
transformed to the ground-state normal mode coordinates. Then the centers of
the displaced Gaussians are found and transformed back to Cartesian
coordinates for an on-the-fly \textit{ab initio} propagation. In addition to
the central trajectory, each polarization of the electric field requires, in
general, two different displaced trajectories to be evaluated.

\section{\label{sec:compdet}Computational details}

All \emph{ab initio} calculations were performed using B3LYP/6-31+G(d,p)
density functional theory for the ground state and time-dependent density
functional theory for the excited state, as implemented in the Gaussian09
package \cite{g09}. The gradient of the transition dipole moment was computed
using a finite-difference approach, as reported in
Ref.~\citenum{Patoz_Vanicek:2018}. The ground-state potential energy surface
was approximated with a global harmonic potential in order to obtain the
initial vibrational state.

The orientational averaging of the 3TGA spectra requires additional
trajectories because the gradient of the transition dipole moment depends on
the orientation of the molecule. In general, one has to compute the spectrum
for three orthogonal orientations of the molecule with respect to the electric
field, which implies one central trajectory for the Condon contribution and
two additional trajectories for each orientation to represent the
Herzberg--Teller contribution to the wavepacket; seven trajectories in total.
In phenyl radical, all seven trajectories have to be evaluated, while in
benzene, since the transition is symmetry-forbidden and since the gradient of
the $z$ component of the transition dipole moment is zero, only four
trajectories are needed (two for $x$ and two for $y$ polarization). In
contrast, the orientational averaging of the ETGA\ spectra requires only a
single \emph{ab initio} trajectory because the averaging can be performed by
changing only the polynomial part of the wavepacket, which depends on the
polarization of the electric field, but does not need additional \emph{ab
initio} data.

A time step of 8 a.u. ($\approx0.194$ fs) was chosen for all trajectories;
3000 steps (giving a total time of 580 fs) were run for the phenyl radical and
5000 steps (970 fs) for benzene. The Hessians of the potential were computed
every four steps, while the intermediate Hessians were obtained by
second-order interpolation.

Gaussian broadening of the resulting spectra was used for the phenyl radical
(half-width at half-maximum of 100 cm$^{-1}$), while for benzene the
autocorrelation function is multiplied by $\cos^{2}(\pi t/2T)$ ($T$ is the
length of the simulation) because this function preserves most of the
autocorrelation function and introduces only slight broadening. However,
longer propagation would be required to simulate the spectrum with peaks as
narrow as in the experiment. For comparison with experiment, unless otherwise
stated, we scale and shift the absorption spectra according to the highest
peak: for phenyl radical, all spectra are shifted by $-437$ cm$^{-1}$, whereas
for benzene we introduce different shifts for the on-the-fly ($3010$ cm$^{-1}%
$), adiabatic harmonic ($3020$ cm$^{-1}$), and vertical harmonic spectra
($3300$ cm$^{-1}$).

\section{\label{sec:resanddisc}Results and discussion}

\subsection{\label{subsec:phenrad}Absorption spectrum of the phenyl radical}

\begin{figure}
[t]\centering\includegraphics[scale=1]{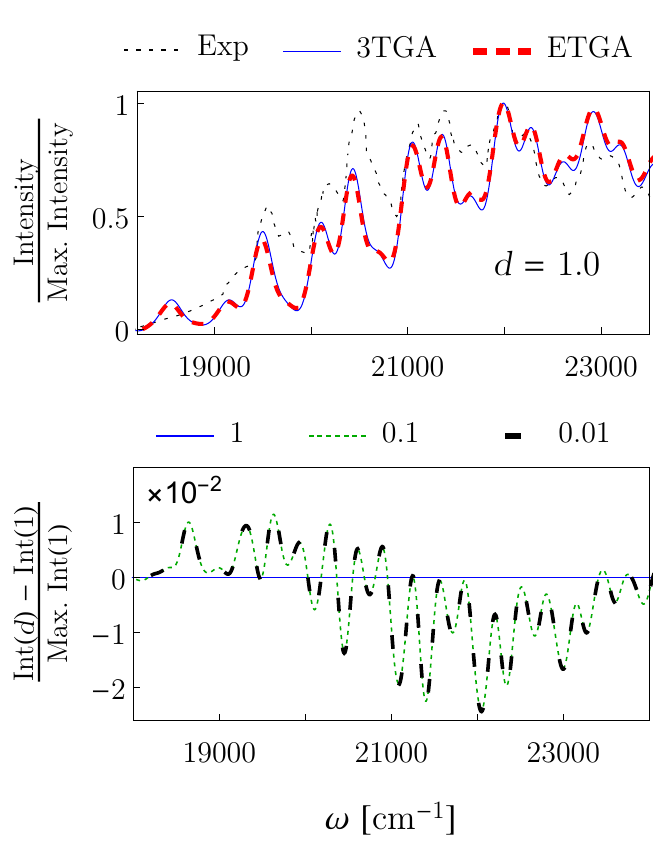}
\caption{\label{fig:PhenylRadical_3G} Orientationally averaged absorption spectra of the phenyl radical calculated
with the 3TGA (using the default value $d = 1$ of the displacement parameter) is compared with the ETGA and
experimental \cite{Radziszewski:1999} spectra in the top panel, and with the 3TGA spectra using smaller values of
the displacement parameter ($d = 0.1$ and $d = 0.01$) in the bottom panel. Since the 3TGA spectra mostly overlap,
the bottom panel shows the scaled difference from the result for $d=1$.}
\end{figure}

The absorption spectrum of the $\tilde{\text{A}}^{2}\text{B}_{1}%
\leftarrow\tilde{\text{X}}^{2}\text{A}_{1}$ electronic transition of the
phenyl radical has been a subject of theoretical investigation
\cite{Kim_Sheng:2002,Biczysko_Barone:2009,Baiardi_Barone:2013,
Patoz_Vanicek:2018} due to a rich vibronic structure originating from the
differences between the ground- and excited-state potential energy surfaces.
Whereas the adiabatic harmonic approach, in which the potential is expanded
about the excited-state minimum, describes the main features of the absorption
spectrum, the vertical harmonic model, which expands the potential about the
ground-state minimum, fails completely, as shown in our previous work
\cite{Patoz_Vanicek:2018}. Here we compute the phenyl radical absorption
spectrum with the on-the-fly \emph{ab initio} 3TGA with different values of
the displacement parameter $d$ and compare the results to the extended TGA
spectrum. Since the anharmonicity of the excited-state potential of the phenyl
radical is shown to influence the spectrum more than the Herzberg--Teller
contribution \cite{Patoz_Vanicek:2018}, this system is suited for further
investigation using the 3TGA, which is specifically intended for treating the
anharmonicity of the Herzberg--Teller active modes. The Herzberg--Teller
component of the wavepacket has a larger spread in position than does the
Condon component, and therefore is more likely to be influenced strongly by
the anharmonicity of the potential. Nevertheless, the calculated 3TGA spectra
(see Fig.~\ref{fig:PhenylRadical_3G}) based on $7$ independent trajectories
overlap almost perfectly with the ETGA result based on a single trajectory.

This is somewhat surprising considering that several Herzberg--Teller modes
are also displaced and, therefore, experience significant anharmonicity of the
potential. The results imply that the local harmonic approximation around the
true center of the wavepacket is valid despite the additional spread and the
change in the shape of the wavepacket. Interestingly, the spectra evaluated
with the 3TGA show almost no dependence on the choice of the initial
displacement parameter $d$. This is encouraging since a strong dependence
would render the 3TGA method impractical; the fact that the dependence is not
only weak but almost nonexistent confirms the results of the single-trajectory
ETGA in the phenyl radical.

\subsection{\label{subsec:benzene}Absorption spectrum of benzene}

The symmetry-forbidden $\text{B}_{2\text{u}}\leftarrow\text{A}_{1\text{g}}$
transition of benzene provides a beautiful example of a Herzberg--Teller
spectrum that is mentioned in many textbooks due to its simple qualitative
interpretation \cite{Herzberg:1966,Hollas:2004,Quack_Merkt:2011,Bernath:2005}.
The experimental and theoretical work on the first excited electronic state of
benzene \cite{Sobolewski_Domcke:1991, Sobolewski_Domcke:1993} and on the
vibronic structure of the corresponding electronic spectrum is so extensive
that we can only refer to a small fraction here
\cite{Sponer_Teller:1939,Atkinson_Parmenter:1978,Fischer_Jakobson:1979,Faulkner_Richardson:1979,Fischer_Knight:1992,Trost_Platt:1997,Etzkorn:1999,Bernhardsson_Serrano-Andres:2000,He_Pollak:2001,Borges_Bielschowsky:2003,Schmied_Scoles:2004,Worth:2007,Loginov_Drabbels:2008,Fally_Vandaele:2009,Penfold_Worth:2009,Li_Lin:2010,Crespo-Otero_Barbatti:2012}%
. The main progression corresponds to a totally symmetric ring-breathing
vibration \cite{Herzberg:1966}. Although this transition is forbidden by
symmetry, its observation in the absorption spectrum is attributed to the
non-totally symmetric vibrations which transform as the $e_{2g}$ irreducible
representation. Li et al. \cite{Li_Lin:2010} include the undisplaced distorted
modes in the calculation of the absorption spectrum using a global harmonic
model without Duschinsky rotation, i.e., by assuming that the ground- and
excited-state normal modes are the same. To go beyond the global harmonic
model, Penfold and Worth \cite{Worth:2007,Penfold_Worth:2009} combine the
construction of an anharmonic potential with the multiconfigurational
time-dependent Hartree wavepacket dynamics, producing excellent results, but
at the cost of a rather detailed inspection of the system required for
reducing the computational cost.

\begin{figure}
[pth]%
\centering\includegraphics[scale=1]{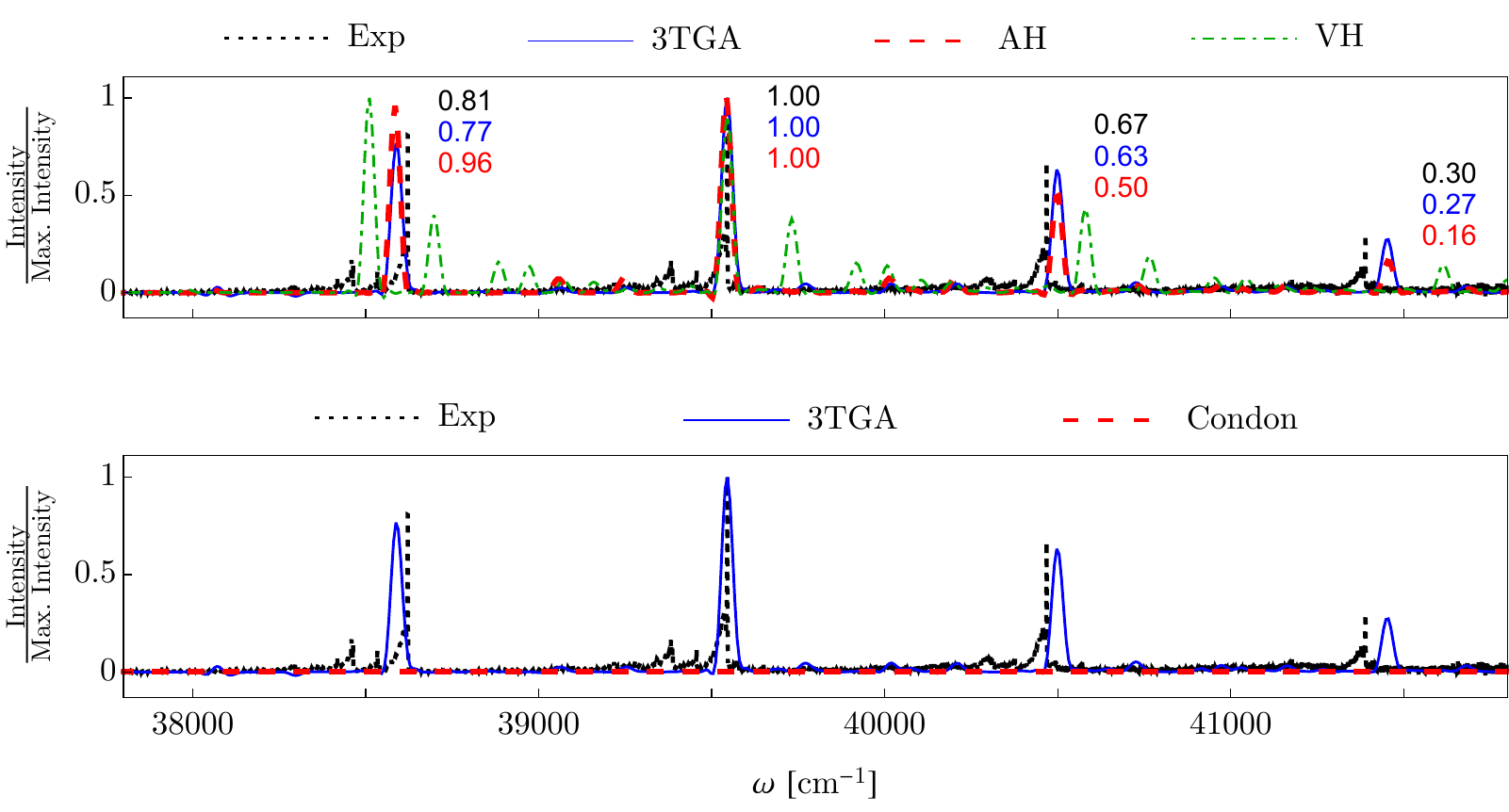}\caption{Orientationally averaged absorption spectrum of benzene calculated with the 3TGA (with $d=1$) is
compared with the
experimental \cite{Fally_Vandaele:2009,Keller-Rudek_Sorensen:2013}
and global harmonic spectra computed using the adiabatic harmonic (AH) and vertical harmonic (VH) models
(top panel). In the bottom panel the full Herzberg--Teller spectrum computed with the 3TGA is compared to the
purely Condon spectrum, which is exactly zero due to symmetry of benzene, but which could be computed with
the original TGA using a single thawed Gaussian.}\label{fig:Benzene_3G_harm_CA}%

\end{figure}

Recently, the on-the-fly \emph{ab initio} ETGA method has been applied to
evaluate the absorption spectrum of benzene with rather high accuracy
\cite{Patoz_Vanicek:2018}. As explained above, this method is an automated,
simple to use single-trajectory method, and the fact that the ETGA was much
more accurate than both vertical and adiabatic global harmonic models is
encouraging. Analogous results for the 3TGA are, therefore, presented in
Fig.~\ref{fig:Benzene_3G_harm_CA}, which compares the on-the-fly \emph{ab
initio} result with the global harmonic spectra (top panel). Note that the
failure of the vertical harmonic model in the phenyl radical and benzene is
not a rule---there are many cases, such as the absorption and photoelectron
spectra of ammonia
\cite{Domcke_VonNiessen:1977,Wehrle_Vanicek:2014,Wehrle_Vanicek:2015}, in
which the vertical is more accurate than the adiabatic harmonic model. Indeed,
the vertical harmonic model describes the Franck-Condon region better and so
might be expected to be a good model for vertical transitions. The on-the-fly
\emph{ab initio} approach, however, overcomes the issue of choosing the
geometry for expanding the potential and is expected to be always at least as
good as the better of the two global harmonic models. Moreover, as
Fig.~\ref{fig:Benzene_3G_harm_CA} shows, the inclusion of anharmonicity in
benzene is essential for obtaining a quantitative agreement---compare the 3TGA
with the adiabatic harmonic model, which is at least qualitatively correct here.

Finally, the bottom panel of Fig.~\ref{fig:Benzene_3G_harm_CA} demonstrates
that in benzene the Herzberg--Teller term, captured with the 3TGA, is
responsible for the observation of the spectrum because the Condon
approximation yields a zero spectrum. This is, indeed, one of the main reasons
why we have implemented the 3TGA; the original TGA is often sufficient for
Condon spectra.

\begin{figure}
[t]%
\centering\includegraphics[scale=1]{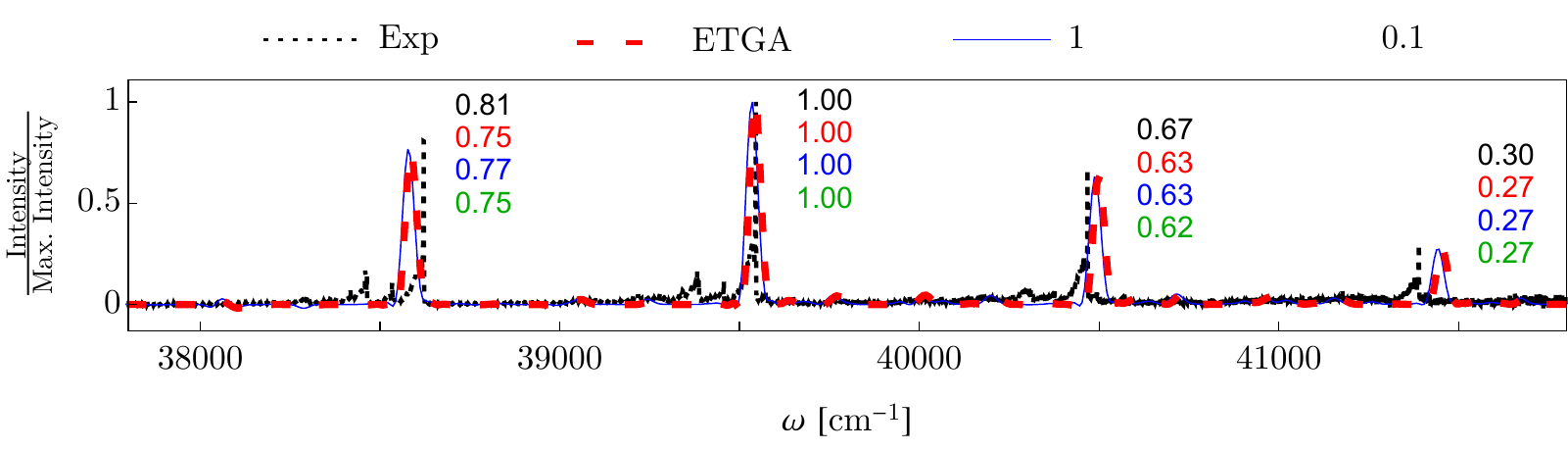}\caption{Orientationally averaged absorption spectra of benzene calculated with the 3TGA for a larger ($d = 1$) and
smaller ($d = 0.1$) displacements are compared with the ETGA and
experimental \cite{Fally_Vandaele:2009,Keller-Rudek_Sorensen:2013} spectra.}\label{fig:Benzene_3G}%

\end{figure}

To investigate the importance of anharmonicity in more detail, the spectra
calculated with the 3TGA and ETGA are compared in Fig.~\ref{fig:Benzene_3G}.
Whereas the 3TGA method with the smaller displacement parameter $d$ gives
almost the same result as the ETGA approach, the 3TGA spectrum calculated with
$d=1$ is red-shifted by $\approx10$ cm$^{-1}$ and the intensity of the first
stronger peak is slightly improved. Due to symmetry, the Herzberg--Teller
modes cannot be displaced from the minimum, so they do not contribute
significantly to the shape of the spectrum. Nevertheless, these modes can
influence the total energy shift of the spectrum, which is in accord with the
observed results.

\subsection{\label{subsec:normfid}Norm conservation and fidelity}

\begin{figure}
[t]%
\centering\includegraphics[scale=1]{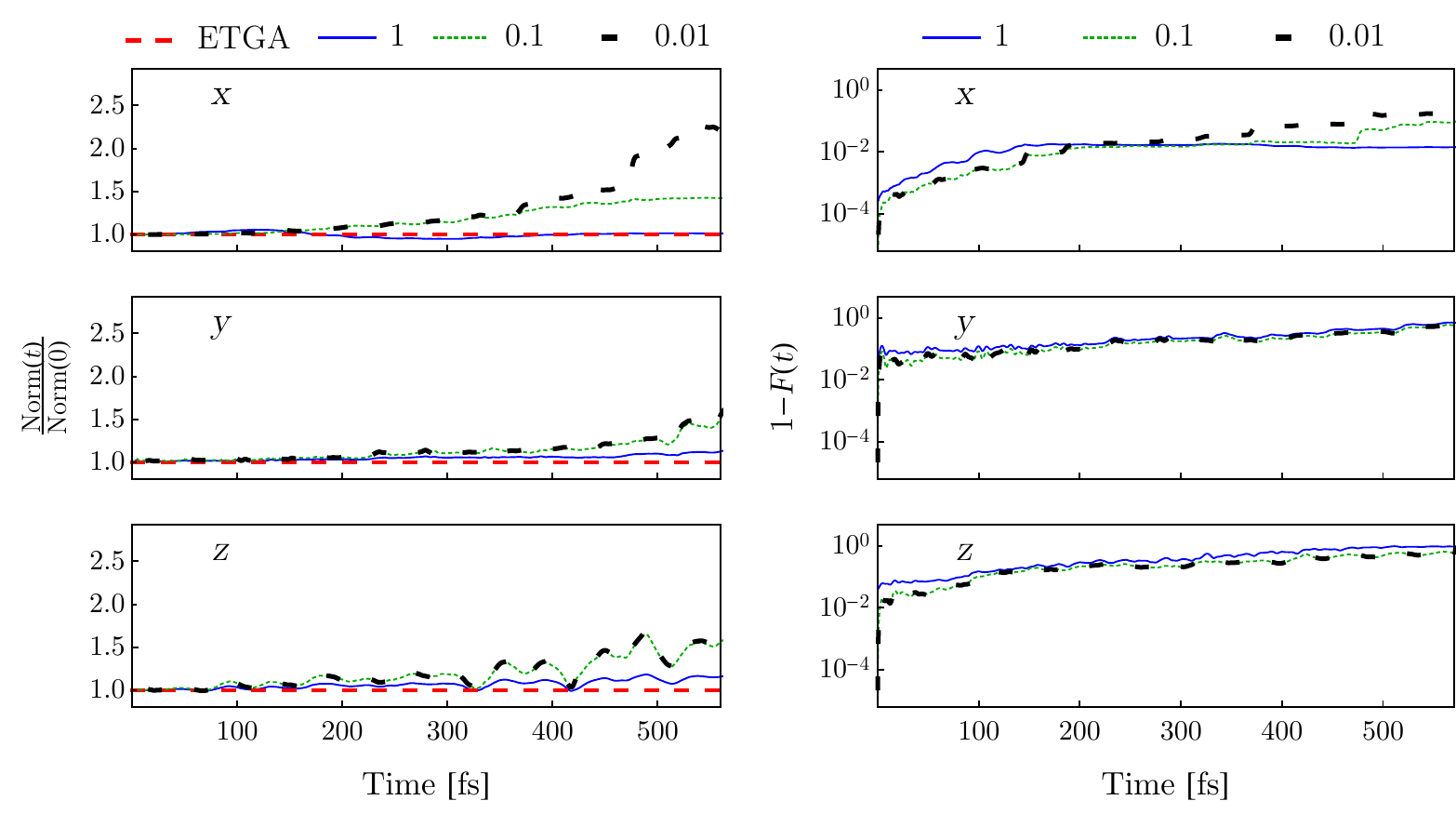}\caption{Norm
(left panels) of the ETGA and 3TGA wavepackets, and fidelity (right panels)
between the ETGA and 3TGA wavepackets [see Eq.~(\ref{eq:fidelity_3G})] for the
phenyl radical. Results for three different values of the displacement parameter ($d =
0.01$, $0.1$, and $1$) as well as for all three different polarizations
($x$, $y$, and $z$) of the electric field are presented.}\label{fig:PhenRadNorm}%

\end{figure}

\begin{figure}
[t]%
\centering\includegraphics[scale=1]{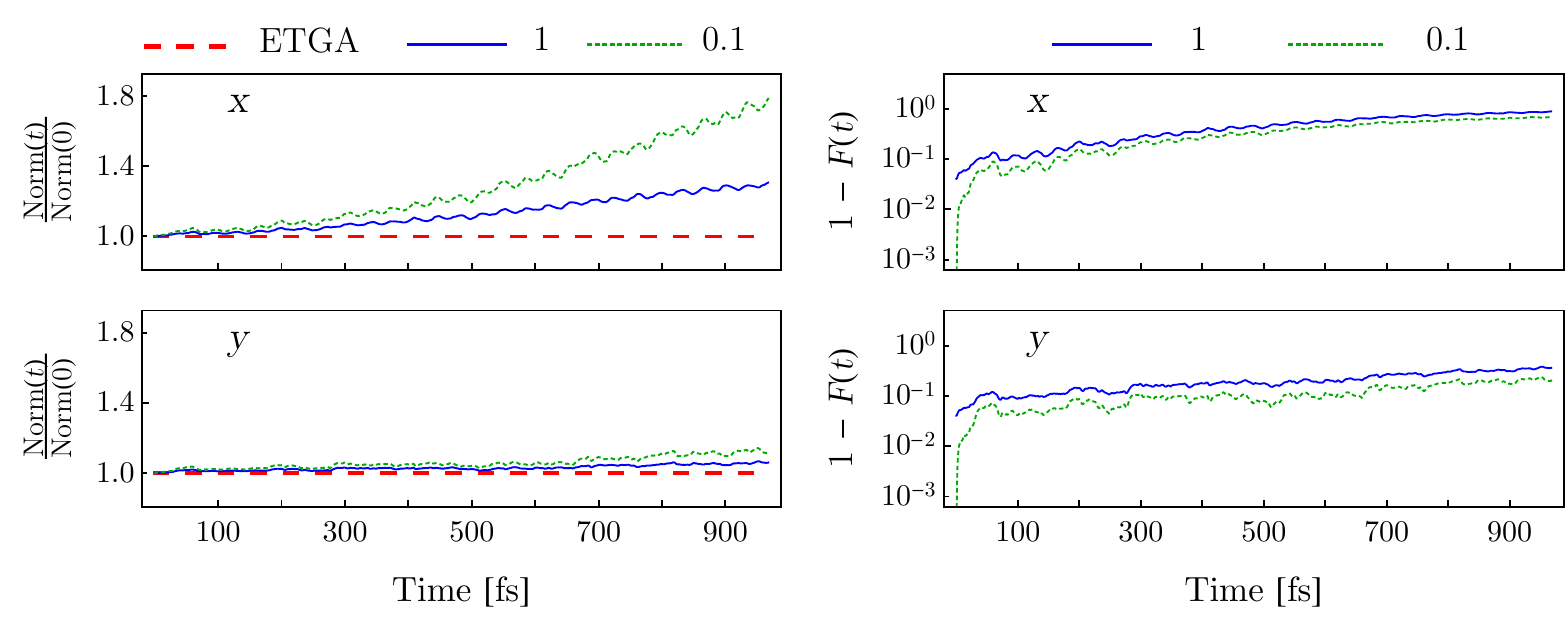}\caption{Norm (left
panels) of the ETGA and 3TGA wavepackets, and the fidelity (right panels) between
the ETGA and 3TGA wavepackets [see Eq.~(\ref{eq:fidelity_3G})] for benzene.
Results for two different values of the displacement parameter ($d = 0.1$ and $1$) and
two different polarizations ($x$ and $y$) of the electric field are
presented. Results for the $z$ polarization are not shown because the corresponding components of the transition dipole moment and its gradient are zero.}\label{fig:BenzeneNorm}
\end{figure}

In contrast with the single-trajectory ETGA, the norm of the 3TGA wavepacket
is not conserved because the 3TGA wavepacket is a superposition of
wavepackets, each of which feels its own time-dependent potential. Indeed, the
numerical results for the phenyl radical (see Fig.~\ref{fig:PhenRadNorm}) and
benzene (see Fig.~\ref{fig:BenzeneNorm}) confirm this theoretical prediction
from Subsection~\ref{subsec:3TGA}. Interestingly, larger displacement ($d=1$)
, which corresponds to placing the additional Gaussians at the extrema of the
initial wavepacket, gives smaller deviations of the norm from unity. In
Eq.~(\ref{eq:3GNorm}) the time-dependent terms are multiplied by the scaling
factor $f_{d}$, indicating that the norm should vary more for smaller
displacements, since the factor $f_{d}$ decreases with the displacement. This
is in contrast with the expectation that the wavepacket propagation, including
the conservation of the norm, should converge to the ETGA with smaller
displacements. The fidelity between the 3TGA and ETGA wavepackets,
\begin{equation}
F(t):=\frac{|\langle\phi^{\text{ETGA}}(t)|\psi^{\text{3TGA}}(t)\rangle|^{2}%
}{||\psi^{\text{ETGA}}(t)||^{2}\,||\psi^{\text{3TGA}}(t)||^{2}}=\frac
{|\langle\phi^{\text{ETGA}}(t)|\psi^{\text{3TGA}}(t)\rangle|^{2}}%
{||\psi^{\text{3TGA}}(t)||^{2}}\,, \label{eq:fidelity_3G}%
\end{equation}
can be used as a measure for comparing the quantum dynamics obtained with the
two semiclassical approximations; it shows deviations of the three thawed
Gaussians from the extended thawed Gaussian wavepacket propagated on the
excited-state potential energy surface. As can be seen in
Figs.~\ref{fig:PhenRadNorm} and \ref{fig:BenzeneNorm}, the fidelity decays
similarly for all values of the displacement parameter, despite the
differences in the initial overlaps. Although the fidelity decays quickly over
time, the final spectra evaluated with the two approximations are nearly the same.

\section{\label{sec:conclusion}Conclusion}

We have presented the 3TGA, constructed by replacing the Herzberg--Teller part
of the initial wavepacket with two displaced Gaussians, in order to describe
electronic spectra beyond the Condon approximation and anharmonicity effects
beyond the single-trajectory ETGA.

The 3TGA presented here is still a very rough approximation to the exact
nuclear wavepacket dynamics. Nevertheless, compared to the global harmonic
models, which are frequently employed in computational chemistry, the
on-the-fly \emph{ab initio} methods based on the TGA offer a rather
computationally cheap way to partially include anharmonicity. Moreover, the
3TGA allows us to analyze the validity of the single-trajectory ETGA; our
results confirm that the additional spread of the wavepacket, induced by the
Herzberg--Teller contribution, does not lead to a significant wavepacket
splitting during the dynamics induced by electronic absorption in the phenyl
radical and benzene.

To conclude, the on-the-fly ab initio implementation of the 3TGA is capable of
describing anharmonicity effects and Herzberg--Teller contribution to the
spectra, as well as of evaluating both symmetry-forbidden and weakly allowed
spectra. The extension beyond the single-trajectory ETGA approach suggests the
viability of simple semiclassical methods that can include wavepacket
splitting and associated interference, and which would readily outperform
global harmonic and single-trajectory methods, while remaining computationally
feasible compared to the more advanced semiclassical and quantum methods.
Indeed, a related pragmatic approach of using a small number of well-defined,
rather than sampled initial conditions, was employed in the multiple coherent
states time-averaged semiclassical initial value representation (MC-TA-SC-IVR)
\cite{Gabas_Ceotto:2017} and its \textquotedblleft
divide-and-conquer\textquotedblright\ extension \cite{Ceotto_Conte:2017},
which were used to evaluate positions of peaks in vibrational spectra with
rather high accuracy. Another appealing feature of the 3TGA is that, despite
its simplicity, the resulting spectrum contains the information not only about
the positions but also about the intensities of the peaks; this is because the
initial conditions are determined by the shape of the initial wavepacket. Last
but not least, like MC-TA-SC-IVR, but in contrast to some semiclassical
methods that require tens of thousands of trajectories for convergence, the
3TGA, by using only three uniquely defined trajectories (we recommend using
$d=1$ always) does not destroy the appealing intuitive picture provided by the
original or extended TGA based on a single trajectory. As a result, we
anticipate further development of efficient methods using only a few
trajectories for semiclassical propagation.

\section*{Acknowledgments}

The authors acknowledge the financial support from the European Research
Council (ERC) under the European Union's Horizon 2020 research and innovation
programme (grant agreement No. 683069 -- MOLEQULE), from the Swiss National
Science Foundation through the NCCR MUST (Molecular Ultrafast Science and
Technology) Network, and from the COST action MOLIM (Molecules in Motion). The
authors are grateful to Frank Grossmann, David Tannor, and Marius Wehrle for discussions.

\appendix

\section{\label{sec_app:orient_av}Orientational average}

Let $\overleftrightarrow{T}$ be a second-order tensor in $3$ dimensions (i.e.,
a $3\times3$ matrix) and let us evaluate the average of the scalar
\begin{equation}
T(\vec{\epsilon}):=\vec{\epsilon}^{T}\cdot\overleftrightarrow{T}\cdot
\vec{\epsilon} \label{eq:tensor_proj}%
\end{equation}
over all unit vectors $\vec{\epsilon}$, i.e.,%
\begin{equation}
\overline{T(\vec{\epsilon})}:=\frac{1}{4\pi}\int_{S^{2}}T(\vec{\epsilon
})d\Omega. \label{eq:tensor_or_avg_def}%
\end{equation}
($\int_{S^{2}}\cdots d\Omega$ denotes the integration over the two-dimensional
unit sphere $S^{2}$.) We shall prove that the result of this average is simply
one third of the trace of the tensor:%
\begin{equation}
\overline{T(\vec{\epsilon})}=\frac{1}{3}\operatorname*{Tr}%
\overleftrightarrow{T}. \label{eq:tensor_or_avg_trace}%
\end{equation}
In particular, in Cartesian coordinates,%
\begin{equation}
\overline{T(\vec{\epsilon})}=\frac{1}{3}(T_{xx}+T_{yy}+T_{zz}).
\label{eq:tensor_or_avg_cartesian}%
\end{equation}

\textbf{Proof}: In polar coordinates $(\theta,\varphi)$, the unit vector
$\vec{\epsilon}$ is expressed as
\begin{equation}
\vec{\epsilon}=\left(  \sin{\theta}\cos{\varphi},\ \sin{\theta}\sin{\varphi
},\ \cos{\theta}\right)  . \label{eq:epsilon}%
\end{equation}
Using Eq.~(\ref{eq:epsilon}), scalar~(\ref{eq:tensor_proj}) becomes%
\begin{align*}
T(\vec{\epsilon})  &  =\left(  T_{xx}\cos^{2}\varphi+T_{yy}\sin^{2}%
\varphi\right)  \sin^{2}\theta+T_{zz}\cos^{2}\theta\\
&  +\left(  T_{xy}+T_{yx}\right)  \sin^{2}\theta\sin\varphi\cos\varphi+\cdots,
\end{align*}
where $\cdots$ denote additional, analogous cross terms for $T_{xz}+T_{zx}$
and $T_{yz}+T_{zy}$. The integration over the unit sphere,%
\[
\overline{T(\vec{\epsilon})}=\frac{1}{4\pi}\int_{0}^{\pi}d\theta\sin\theta
\int_{0}^{2\pi}d\varphi\,T(\vec{\epsilon}),
\]
suppresses all the cross terms; e.g., $\sin\varphi\cos\varphi=(1/2)\sin\left(
2\varphi\right)  $ and the integration over $\varphi$ of the cross term for
$T_{xy}+T_{yx}$ is, therefore, zero. Integration of the diagonal terms over
$\varphi$ and making a substitution $u:=\cos\theta$ gives%
\begin{align*}
\overline{T(\vec{\epsilon})}  &  =\frac{1}{4}\int_{-1}^{1}du\left[  \left(
T_{xx}+T_{yy}\right)  (1-u^{2})+2T_{zz}u^{2}\right] \\
&  =\frac{1}{4}\left[  \left(  T_{xx}+T_{yy}\right)  (u-u^{3}/3)+2T_{zz}%
u^{3}/3\right]  _{-1}^{1}\\
&  =\frac{1}{3}\left(  T_{xx}+T_{yy}+T_{zz}\right)  =\frac{1}{3}%
\operatorname*{Tr}\overleftrightarrow{T},
\end{align*}
where in the last step we used the invariance of the trace of a tensor under
coordinate transformations, in order to obtain the coordinate-independent
expression~(\ref{eq:tensor_or_avg_trace}).

\section{\label{sec_app:grad_mu_NAC}Proof of the expression (\ref{eq:grad_mu})
for the gradient of the transition dipole moment}

Before it is expressed in the basis of electronic states, the molecular
electric dipole operator is simply a sum
\[
\hat{\vec{\mu}}_{\text{mol}}\equiv\vec{\mu}_{\text{mol}}(q,\{\hat{\vec{r}}%
_{n}\})=\vec{\mu}_{\text{nu}}(q)+\vec{\mu}_{\text{el}}(\{\hat{\vec{r}}_{n}\})
\]
of its nuclear component $\vec{\mu}_{\text{nu}}(q)$ given in
Eq.~(\ref{eq:mu_nu}) and electronic component%
\begin{equation}
\vec{\mu}_{\text{el}}(\{\hat{\vec{r}}_{n}\})=-e\sum_{n=1}^{N_{\text{el}}}%
\hat{\vec{r}}_{n}, \label{eq:mu_el}%
\end{equation}
where $\vec{r}_{n}$ are the Cartesian coordinates of the $n$th electron and
$\{\vec{r}_{n}\}:=(\vec{r}_{1},\ldots,\vec{r}_{N_{\text{el}}})$. Relation
(\ref{eq:grad_mu}) is proven directly by taking the gradient of
Eq.~(\ref{eq:mu_ab}):
\begin{align}
\partial_{j}\vec{\mu}_{\alpha\beta}(q)  &  =\langle\partial_{j}\alpha
|\hat{\vec{\mu}}_{\text{mol}}|\beta\rangle+\langle\alpha|\hat{\vec{\mu}%
}_{\text{mol}}|\partial_{j}\beta\rangle+\langle\alpha|\partial_{j}\hat
{\vec{\mu}}_{\text{mol}}|\beta\rangle\label{eq:grad_mu_components}\\
&  =\sum_{\gamma}\left(  \langle\partial_{j}\alpha|\gamma\rangle\langle
\gamma|\hat{\vec{\mu}}_{\text{mol}}|\beta\rangle+\langle\alpha|\hat{\vec{\mu}%
}_{\text{mol}}|\gamma\rangle\langle\gamma|\partial_{j}\beta\rangle\right)
+\langle\alpha|\partial_{j}\vec{\mu}_{\text{nu}}(q)|\beta\rangle\\
&  =\sum_{\gamma}\left(  F_{\alpha\gamma,j}^{\dag}\vec{\mu}_{\gamma\beta}%
+\vec{\mu}_{\alpha\gamma}F_{\gamma\beta,j}\right)  +\partial_{j}\vec{\mu
}_{\text{nu}}(q)\delta_{\alpha\beta}\\
&  =\sum_{\gamma}\left(  -F_{\alpha\gamma,j}\vec{\mu}_{\gamma\beta}+\vec{\mu
}_{\alpha\gamma}F_{\gamma\beta,j}\right)  +\partial_{j}\vec{\mu}_{\text{nu}%
}(q)\delta_{\alpha\beta}\\
&  =\left[  \vec{\bm{\mu}},\mathbf{F}_{j}\right]  _{\alpha\beta}+\partial
_{j}\vec{\mu}_{\text{nu}}(q)\delta_{\alpha\beta}.
\end{align}
The first equality follows from the Leibniz law; the second equality employs a
resolution of identity over electronic states and takes into account that the
electronic dipole moment operator (\ref{eq:mu_el}) is independent of nuclear
coordinates; the third equality follows from the definition (\ref{eq:F_ab_j})
of nonadiabatic coupling vectors, the fact that $\partial_{j}\vec{\mu
}_{\text{nu}}(q)$ is a purely nuclear operator, and orthogonality of the
electronic states; the fourth equality uses the antihermitian property of the
matrix of nonadiabatic couplings,%
\[
\mathbf{F}_{j}^{\dag}=-\mathbf{F}_{j},
\]
which follows from the orthogonality of the electronic states,%
\[
0=\partial_{j}\delta_{\alpha\beta}=\partial_{j}\langle\alpha(q)|\beta
(q)\rangle=F_{\alpha\beta,j}^{\dag}+F_{\alpha\beta,j};
\]
the fifth step completes the proof.

\section{\label{sec_app:A_gamma_prop}Propagation of the width and phase of the
thawed Gaussian wavepacket}

The approach by Lee and Heller \cite{Lee_Heller:1982} suggests splitting the
complex symmetric width matrix $A$ into a product of two matrices $P$ and
$Z$:
\begin{equation}
A_{t}=-\frac{i}{2\hbar}P_{t}\cdot Z_{t}^{-1}\,. \label{eq:A_t}%
\end{equation}
By imposing $\dot{Z}_{t}=m^{-1}\cdot P_{t}$, with the initial conditions
$Z_{0}=I$, where $I$ is the identity matrix, and $P_{0}=2i\hbar A_{0}$, the
expression for the propagation of $Z$ and $P$ parameters is obtained:
\begin{equation}%
\begin{pmatrix}
Z_{t}\\
P_{t}%
\end{pmatrix}
=M_{t}%
\begin{pmatrix}
Z_{0}\\
P_{0}%
\end{pmatrix}
\,. \label{eq:ZP}%
\end{equation}

Regarding $\gamma_{t}$, which is a generalization of classical action and
represents both the dynamical phase and normalization, it is evaluated as
\begin{align}
\gamma_{t}  &  =\gamma_{0}+\int_{0}^{t}L_{\tau}d\tau-\hbar^{2}\int_{0}%
^{t}\operatorname{Tr}\left(  m^{-1}\cdot A_{\tau}\right)  d\tau\\
&  =\gamma_{0}+\int_{0}^{t}L_{\tau}d\tau-\hbar^{2}\int_{0}^{t}%
\operatorname{Tr}\left(  -\frac{i}{2\hbar}\dot{Z}_{\tau}\cdot Z_{\tau}%
^{-1}\right)  d\tau\\
&  =\gamma_{0}+\int_{0}^{t}L_{\tau}d\tau+\frac{i\hbar}{2}\ln\left(  \det
Z_{t}\right)  \,, \label{eq:gamma_t}%
\end{align}
where the conditions imposed on $Z$ and $P$ are used in order to obtain the
final expression. Note that, since the determinant in the final expression is
complex, a proper branch of the logarithm has to be taken in order to make
$\gamma_{t}$ continuous in time. If continuity were not imposed on $\gamma
_{t}$, the wavepacket would show sudden jumps by $\pi$ in the overall phase.
Phase continuity is also important in the evaluation of the correlation
function, which comprises a square root of a complex determinant $\det
(A_{0}+A_{t}^{\ast})$. The continuity of the correlation function is enforced
by taking the appropriate branch of the square root.

\section{\label{sec_app:ETGA}Derivation of the extended thawed Gaussian
approximation}

\textbf{Proof of Eq.~(\ref{eq:phi_t_p_polynomial}): }The effective potential
(\ref{eq:effPot}) depends on $t$ only implicitly, via $q_{t}$. Considering
$V_{\text{eff}}$ as a function of $q$ and $q_{t}$, its dependence on $q_{t}$
is%
\begin{equation}
\frac{\partial V_{\text{eff}}(q,q_{t})}{\partial q_{t}}=\frac{1}{2}%
V^{\prime\prime\prime}|_{q_{t}}\left(  q-q_{t},q-q_{t},\cdot\right)
\label{eq:dVeff_dqt_exact}%
\end{equation}
because in the derivation of Eq.~(\ref{eq:dVeff_dqt_exact}) the sum
$\partial_{q}V|_{q_{t}}+\operatorname{Hess}_{q}V|_{q_{t}}\cdot(q-q_{t})$
appears twice, with opposite signs, and therefore cancels. [Here
$V^{\prime\prime\prime}$ is a rank-$3$ tensor of the third derivatives of $V$
and the dot $\cdot$ in the third argument of $V^{\prime\prime\prime}$
indicates that the tensor has been only partially contracted; the right-hand
side of Eq.~(\ref{eq:dVeff_dqt_exact}) is still a vector.] Within local
harmonic approximation (LHA), where all derivatives of $V$ at $q_{t}$ beyond
the second are neglected,%
\begin{equation}
\frac{\partial V_{\text{eff}}}{\partial q_{t}}\overset{\text{LHA}}{=}0\text{.}
\label{eq:dVeff_dqt}%
\end{equation}

Equation~(\ref{eq:dVeff_dqt}) implies that the effective Hamiltonian operator
(\ref{eq:effHam}), considered as a function of initial conditions $q_{0}$ and
$p_{0}$, $\hat{H}_{\text{eff}}(q_{0},p_{0},t)\equiv H_{\text{eff}}(\hat
{q},\hat{p},q_{0},p_{0},t)$, is independent of $p_{0}$:
\begin{equation}
\frac{\partial\hat{H}_{\text{eff}}}{\partial p_{0}}=\left(  \frac{\partial
q_{t}}{\partial p_{0}}\right)  ^{T}\cdot\frac{\partial\hat{V}_{\text{eff}}%
}{\partial q_{t}}\overset{\text{LHA}}{=}0\text{.} \label{eq:dHeff_dp0}%
\end{equation}
The effective time evolution operator
\[
\hat{U}(q_{0},p_{0},t):=\mathcal{T}\exp\left[  -\frac{i}{\hbar}\int_{0}%
^{t}\hat{H}_{\text{eff}}(q_{0},p_{0},\tau)d\tau\right]
\]
induced by $\hat{H}_{\text{eff}}$ is, in general, a function of $q_{0}$ and
$p_{0}$, but factorizing the time-ordered product into a product of
exponentials for infinitesimal time steps, expanding the exponential for each
time step into a Taylor series, and using Eq.~(\ref{eq:dHeff_dp0}) shows that,
within the LHA,%
\begin{equation}
\frac{\partial\hat{U}(q_{0},p_{0},t)}{\partial p_{0}}\overset{\text{LHA}}{=}0
\label{eq:dU_dp0}%
\end{equation}
and, indeed, any polynomial in derivatives with respect to $p_{0}$ acting on
$\hat{U}$ vanishes:%
\begin{equation}
P\left(  \frac{\hbar}{i}\frac{\partial}{\partial p_{0}}\right)  \hat{U}%
(q_{0},p_{0},t)\overset{\text{LHA}}{=}0. \label{eq:P_on_U}%
\end{equation}

The initial Herzberg-Teller state $|\phi_{0}\rangle$ propagated to time $t$ is%
\begin{align}
|\phi_{t}\rangle &  =\hat{U}(q_{0},p_{0},t)|\phi_{0}\rangle=\hat{U}%
(q_{0},p_{0},t)P\left(  \frac{\hbar}{i}\frac{\partial}{\partial p_{0}}\right)
|\psi_{0}\rangle\nonumber\\
&  =P\left(  \frac{\hbar}{i}\frac{\partial}{\partial p_{0}}\right)  \hat
{U}(q_{0},p_{0},t)|\psi_{0}\rangle=P\left(  \frac{\hbar}{i}\frac{\partial
}{\partial p_{0}}\right)  |\psi_{t}\rangle, \label{eq:phi_t_from_psi_t}%
\end{align}
where the Dirac kets $|\psi_{0}\rangle$ and $|\psi_{t}\rangle$ are considered
as functions of $q_{0}$ and $p_{0},$ and Eq.~(\ref{eq:P_on_U}) was used to
switch the order of $P$ and $\hat{U}$. This completes the proof of
Eq.~(\ref{eq:phi_t_p_polynomial}).

\textbf{Proof of Eqs.~(\ref{eq:phi_t_HTA}) and
(\ref{eq:polynomial_eq_of_motion}): }Let us evaluate the derivative
$\partial|\psi_{t}\rangle/\partial p_{0}$, needed in
Eq.~(\ref{eq:phi_t_from_psi_t}), analytically in position representation, in
which $|\psi_{t}\rangle$ is the TGA wavefunction~(\ref{eq:GWP}):%
\begin{align}
&  \frac{\partial\psi_{t}(q)}{\partial p_{0}}\overset{\text{LHA}}{=}\left(
\frac{\partial q_{t}}{\partial p_{0}}\right)  ^{T}\cdot\frac{\partial\psi_{t}%
}{\partial q_{t}}+\left(  \frac{\partial p_{t}}{\partial p_{0}}\right)
^{T}\cdot\frac{\partial\psi_{t}}{\partial p_{t}}+\left(  \frac{\partial
\gamma_{t}}{\partial p_{0}}\right)  ^{T}\cdot\frac{\partial\psi_{t}}%
{\partial\gamma_{t}}\nonumber\\
&  =\left\{  M_{t,qp}^{T}\cdot\left[  2A_{t}\cdot\left(  q-q_{t}\right)
-\frac{i}{\hbar}p_{t}\right]  +M_{t,pp}^{T}\cdot\frac{i}{\hbar}\left(
q-q_{t}\right)  +M_{t,qp}^{T}\cdot p_{t}\frac{i}{\hbar}\right\}  \psi
_{t}(q),\nonumber\\
&  =\left(  2M_{t,qp}^{T}\cdot A_{t}+\frac{i}{\hbar}M_{t,pp}^{T}\right)
\cdot\left(  q-q_{t}\right)  \psi_{t}(q). \label{eq:dpsi_t_dp0}%
\end{align}
In the first step of the derivation, we neglected the dependence of the width
matrix $A_{t}$ on $p_{0}$, which follows from Eq.~(\ref{eq:A_dot}) within the
LHA. In the second step, we used the derivative%
\[
\frac{\partial\gamma_{t}}{\partial p_{0}}=\frac{\partial S_{t}\left(
q_{0},p_{0}\right)  }{\partial p_{0}}=\left(  \frac{\partial q_{t}}{\partial
p_{0}}\right)  ^{T}\cdot\frac{\partial S_{t}\left(  q_{0},q_{t}\right)
}{\partial q_{t}}=M_{t,qp}^{T}\cdot p_{t},
\]
where $S_{t}=\int_{0}^{t}L_{\tau}d\tau$ is the classical action component of
$\gamma_{t}$ expressed either as a function of $q_{0}$ and $p_{0}$ or of
$q_{0}$ and $q_{t}$. The derivation of the central ETGA equations
(\ref{eq:phi_t_HTA}) and (\ref{eq:polynomial_eq_of_motion}) is completed by
substituting the Herzberg-Teller form [Eq.~(\ref{eq:HTA})] of the polynomial
$P$,
\[
P(x)=\mu_{21}\left(  q_{0}\right)  +\partial_{q}\mu_{21}|_{q_{0}}^{T}\cdot x,
\]
into Eq.~(\ref{eq:phi_t_p_polynomial}), giving%
\[
\phi_{t}(q)=\left[  \mu_{21}\left(  q_{0}\right)  +\partial_{q}\mu
_{21}|_{q_{0}}^{T}\cdot\frac{\hbar}{i}\frac{\partial}{\partial p_{0}}\right]
\psi_{t}(q),
\]
and using expression (\ref{eq:dpsi_t_dp0}) for $\partial\psi_{t}(q)/\partial
p_{0}$.

\section{\label{sec_app:3TGA_extrema}Extrema of the Herzberg--Teller part of
the wavepacket}

The extrema of the wavepacket (\ref{eq:initial_HT}) are found by setting its
gradient to zero. The gradient, given by
\begin{equation}
\partial_{q}\phi_{0}^{\text{ETGA-HT}}(q)=\left[  \partial_{q}\mu-2\left(
\partial_{q}\mu^{T}\cdot\Delta q\right)  A_{0}\cdot\Delta q\right]  g_{q_{0}%
}(q)\,, \label{eq_app:der_phi}%
\end{equation}
where $\Delta q=q-q_{0}$, will be zero if and only if
\begin{equation}
\partial_{q}\mu=2(\partial_{q}\mu^{T}\cdot\Delta q)A_{0}\cdot\Delta q
\label{eq_app:quad_eq}%
\end{equation}
because the Gaussian function $g_{q_{0}}(q)$ is strictly positive. Since the
initial width matrix $A_{0}$ is positive definite, it has an inverse and we
can multiply Eq.~(\ref{eq_app:quad_eq}) on the left with $A_{0}^{-1}$, which
gives%
\begin{equation}
A_{0}^{-1}\cdot\partial_{q}\mu=2(\partial_{q}\mu^{T}\cdot\Delta q)\Delta q
\label{eq_app:quad_eq_proj1}%
\end{equation}
Scalar product of Eq.~(\ref{eq_app:quad_eq_proj1}) with the vector
$\partial_{q}\mu$ gives%
\begin{equation}
\partial_{q}\mu^{T}\cdot A_{0}^{-1}\cdot\partial_{q}\mu=2(\partial_{q}\mu
^{T}\cdot\Delta q)^{2}, \label{eq_app:quad_eq_proj2}%
\end{equation}
with two solutions
\begin{equation}
\partial_{q}\mu^{T}\cdot\Delta q=\pm\sqrt{\frac{1}{2}\partial_{q}\mu^{T}\cdot
A_{0}^{-1}\cdot\partial_{q}\mu}. \label{eq_app:extrema_final_eq}%
\end{equation}
Finally, substitution of $\partial_{q}\mu^{T}\cdot\Delta q$ from
Eq.~(\ref{eq_app:extrema_final_eq}) into Eq.~(\ref{eq_app:quad_eq_proj1})
yields%
\[
\Delta q=\pm\frac{\sqrt{2}}{2}\frac{A_{0}^{-1}\cdot\partial_{q}\mu}%
{\sqrt{\partial_{q}\mu^{T}\cdot A_{0}^{-1}\cdot\partial_{q}\mu}},
\]
where it is easy to see that the positive sign corresponds to the local
maximum and negative sign to the local minimum, completing the proof of
Eqs.~(\ref{eq:max_HT}) and (\ref{eq:Delta_q}).

\bibliographystyle{aipnum4-1}
\bibliography{Herzberg-Teller_3TGA}

\end{document}